# The cost of uncertainty

Carlos Esteban POSADA[1]

Abstract

The traditional and most common view of economists on the issue of (bad) uncertainty and its effects has been one of partial equilibrium. When the topic is approached from a macroeconomic perspective, the most frequent has been the examination of the effects of uncertainty shocks on short-term dynamics with various methods, but mainly with neo-Keynesian and statistical (vector autoregressive) models. This document responds to another concern and has two objectives: 1) to reflect on this issue with some instruments of the macroeconomist´s toolbox related to a medium or long-term horizon, and 2) report a ciphering of the social cost of uncertainty in the Colombian case.

Key Words: Interest rate, Investment, Output growth rate, Risk, Savings, Uncertainty

[1] Professor. School of Finance, Economics and Government; Universidad EAFIT (Colombia). Address: cposad25@eafit.edu.co

## I. Introduction

The traditional and most usual view of economists on the effects of (bad) uncertainty (risk) has been one of partial equilibrium. These effects are analyzed in one market or another, or the issue of their effects on household savings or firm performance is addressed. Moreover, that view has been dominated by short-term concerns[2]. When the subject is studied from a macroeconomic perspective, the most frequent approach has been the measurement of the effects of various uncertainty shocks on short-run dynamics with various methods, but mainly with neo-Keynesian or statistical (autoregressive vectors) models[3].

The statements in the previous paragraph require a comment: there has been an important stream of economic thinking taking into consideration uncertainty from a macroeconomic perspective: the one originated in the inclusion of the uncertainty issue in that of long-run economic growth (Brock and Mirman, 1972 and 1973; a derivation of this is that of "real business cycles"). But this whole large theoretical stream incorporates assumptions leading to the long-run neutrality of the uncertainty (that is, the neutrality of the average value of a series of random shocks affecting output via technology) with respect to the outcomes of the endogenous variables of growth and cycles models.

This paper responds to another concern (and so uses other hypotheses) and has two objectives: a) to reflect on the issue of significant real effects of uncertainty on a long-term horizon, omitting issues of money demand, monetary policy and implications of price and nominal wage rigidities, and b) to report our ciphering of the social cost of uncertainty in the Colombian case, seeking coherence with the first objective. Sections II, III and IV cover these targets. Section V presents the model on which this calculation is based. Section VI reports the main conclusion.

## II. Risk aversion

It is traditional to consider that attitude to risk can be classified into three groups: risk-loving, risk-neutral and risk-averse individuals. Economists have considered the case of risk aversion as the usual one and, given this, have developed the theory of financial asset prices and rates of return (here and in what follows we will consider interest rates and rates of return to be synonymous terms), distinguishing between risk-free assets (default risk; the typical example has been the U.S. government bond) and risky assets (from lower to higher risk)[4], and, hence, the existence of risk premia added to interest rates on risk-free assets. For the sake of simplicity, let us concentrate on two rates: the risk-free rate and the risky rate; the difference in between is the risk premium. Thus, any factor capable of raising the risk premium, given the risk-free rate, will cause an increase in the risky rate[5].

---

[2] In what follows, uncertainty and risk are equivalent, assuming that they are measurable. Frank Knight insisted on the opposite: that uncertainty (measurable and therefore insurable) and risk, conceived by him as unquantifiable, are different (Bloom, 2014).

[3] Notwithstanding there are studies addressing the issue of the medium- or long-term consequences of uncertainty (Bloom 2014 reviews some of these studies). One of these, the most famous, is that of Ramey and Ramey (1995), which reports international econometric evidence of the inverse relationship between economic growth rates and instability. Campbell *et al.* (2020) examine the medium-term effects of variations in the degree of uncertainty on interest rates and share prices related to changes in inflation and output gap.

Notwithstanding, the following case is feasible: an increase in the risk premium (associated, for example, with greater volatility of macroeconomic variables) and, at the same time, such a large fall in the risk-free rate inducing a reduction in the risk-bearing rate despite the increase in the risk premium. In the following section we will discuss this case and the case of increases in the risk premium that go in the same direction as those of the risk-free and risky rates.

## III. Uncertainty in "robust" and "frail" economies

Higher overall uncertainty in the "robust economy" (the reason for this name and quotation marks is clarified below) has an impact in several fields: a) higher precautionary savings of households and a consequent increase of private savings relative to their disposable income (Carroll *et al*., 2019)[6] ; higher preference for security in the choice of financial assets (Tobin, 1958) and, then, for public debt securities to the detriment of the demand for private debt securities and equities; falls in interest rates on public debt securities (without default risk; here and in all that follows reference is made to rates net of inflation expectations, that is, real rates); moreover, investment tends to fall (Caballero and Pindyck, 1996).

Thus, in net terms, there is an excess of savings over investment and, therefore, the average interest rate (the weighted average of interest rates) falls, eliminating this excess and, with this, counteracting the fall in investment induced by the greater uncertainty (Bloom, 2009 and 2014). Additionally, in the face of a permanent increase in uncertainty, the following can be predicted: a possible change in the yield curve of public debt securities (different variation of very short-term rates compared to those of longer terms), an increase in the rate spreads of private debt with respect to the public one (assuming that public debt securities are free from default risk). Finally, the composition of household financial assets shifts towards less risky values in terms of expectations about probabilities of price losses or debtor insolvency.

---

[4] Wickens (2011, cp. 11) has an excellent presentation of the topic of interest rate and risk premium, and its connection with the variance of macroeconomic series.

[5] This clarification is made because even U.S. "Treasury" bonds, despite not having the risk of insolvency, may have a risk premium associated with a probability of loss of their value in the secondary market due to revisions of inflation expectations and conjectures about changes in the policy´s interest rate (see Segal *et al*., 2015; Campbell *et al*., 2020).

[6] For European households the main driver of savings is a precautionary attitude, according to Horioka and Ventura (2024).

When investment and capital recover or increase in response to the reduction in the interest rate, could be, in the future, the same or a higher pace of innovation and hence, perhaps, a biggest per worker product growth rate (Aghion and Howitt, 2009, chap. 5; hereinafter: A&H).

But what is a "robust economy"? This term could be applied to the following: a) small open economies with broad and unavoidable capital controls[7]; b) large economies with highly credible political (and monetary policy) institutions. In these economies, the interest rate responds to changes in savings or investment functions, even if there are surpluses or deficits in the trade and current accounts of their balance of payments, that is, if there are gaps between national saving and domestic investment.

So, in a "robust economy" an increase in the uncertainty degree among households (for some reason other than uncertainty about political institutions), which generates an expansion of precautionary savings, tends to shift the savings curve to the right (see Figure 1). But this does not lead to very large changes in the balance of payments or to an increase in household demand for financial assets denominated in foreign currencies (nor does it lead the financial intermediaries to significantly change the composition of their assets looking for increase the share of external assets). Consequently, the interest rate tends to fall, albeit with lags, because of households' decisions to increase their savings.

Figure 1 illustrates the main macroeconomic effects of an unanticipated but permanent increase in overall uncertainty in a "robust economy", under the assumption of similar uncertainty´s effects on saving and investment.

[7]Notwithstanding, "broad and unavoidable capital controls" usually are transitory instruments because of its social costs and "time inconsistency" problems (see Chang *et al*. 2024). So, the robustness character based on capital controls is something also transitory.

Figure 1

**In the "robust economy" an increase of uncertainty cause a reduction of the interest rate**

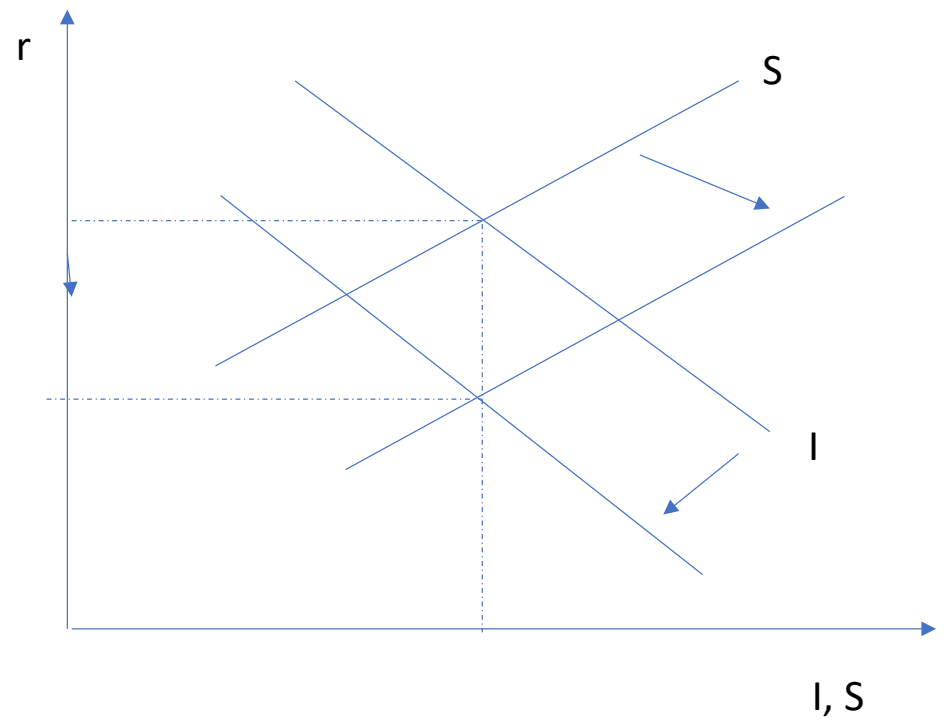

I: investment
S: Saving
r: Real interes rate

Symmetrically, what we called "frail economies" here are more than simply open economies: that is, economies that are also characterized by households having a certain degree of distrust in the soundness of their political institutions (including those of monetary policy).

Thus, in an "frail economy", greater uncertainty also increases the household savings rate, but the bulk of these higher savings is used to purchase foreign currencies and other financial and real assets abroad (a case of "capital flight"), that is, it gives rise to net capital outflows (or lower net capital inflows). So, these facts and reactions usually induces a higher local interest rate and hence a lower investment. The consequent reductions in the capital stock affect negatively the economic activity (Calvo, 1998; Fernández-Villaverde *et al.*, 2011) and, eventually, depress the process of innovation and technical change. Figure 2 illustrates the effects, in this case, on two fundamental variables: investment (= national saving + external saving) and interest rate.

Figure 2

**A higher uncertainty in the "frail economy" causes a higher interest rate, and a lower investment**

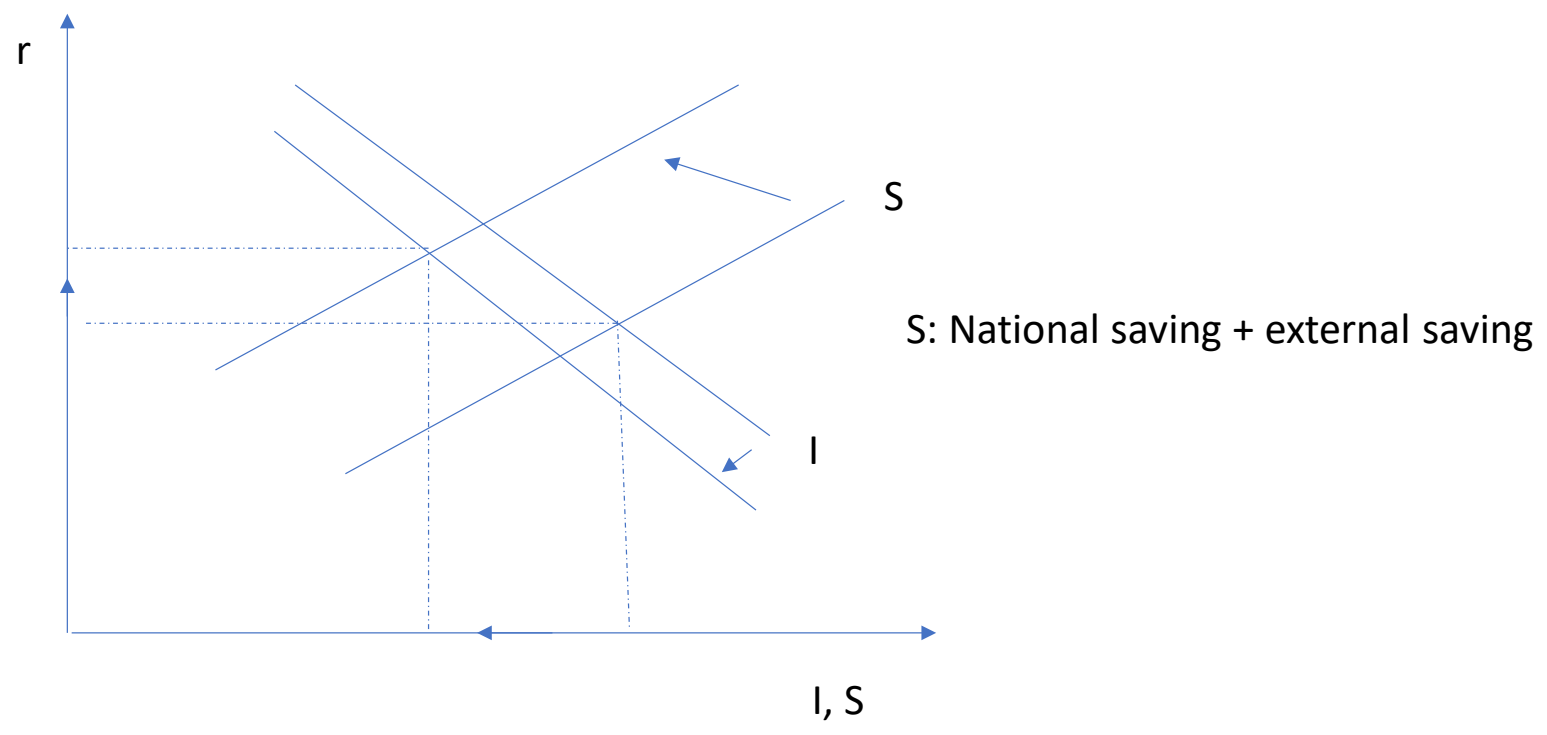

The case of increased precautionary savings in the "frail economy" illustrated in Figure 2 is a case of the loss of relevance of a common assumption among the builders and users of the (now conventional) "small open economy" model, according to which the interest rate is equal to an external rate plus a margin depending positively on the degree of external debt[8]. Indeed, Figure 2 also illustrates a case for a small but frail open economy having an interest rate increase because of higher uncertainty and thus higher precautionary savings, which implies a reduction in external net indebtedness.

## IV. An estimate of the cost of uncertainty

How costly for society is an uncertainty increase in the most relevant case, that of an "frail economy"?

[8] See, for example, Uribe and Schmitt-Grohé, 2017, p. 245-6.

The answer is not easy; It depends on many considerations. Several of them have to do with situations of too great instability, even chaotic, making estimates of the social cost difficult and uncertain, even if we were convinced it is significant. But it is not difficult to estimate it under simplifying assumptions, among which we can include trajectories of economic activity in alternative "steady state" situations, comparing them in the following way: one with a lower and constant uncertainty degree and another with a higher (and also constant) one.

In the following paragraphs I show the results of a ciphering of the social cost for the Colombian case under very simplifying assumptions, which, as mentioned above, must be interpreted as the lower limit of a cost that could be higher under a more complex hypothesis´s set.

For the Colombian case, a practical way to measure a higher uncertainty is to consider what has happened with the ratio between the so-called *EMBI-Colombia and EMBI-Chile*[9]. Graph 1 shows this.

[9] Emerging Market Bonds Index. This index is calculated by *J. P. Morgan Chase Bank*. For each emerging economy, it considers the difference between the yield rate corresponding to a bond issued in dollars by its government and the rate of a bond issued by the United States´s government, for similar terms. The figures on this indicator used here are from the following source: https://www.invenomica.com.ar/riesgo-pais-embi-america-latina-serie-historica/

Graph 1

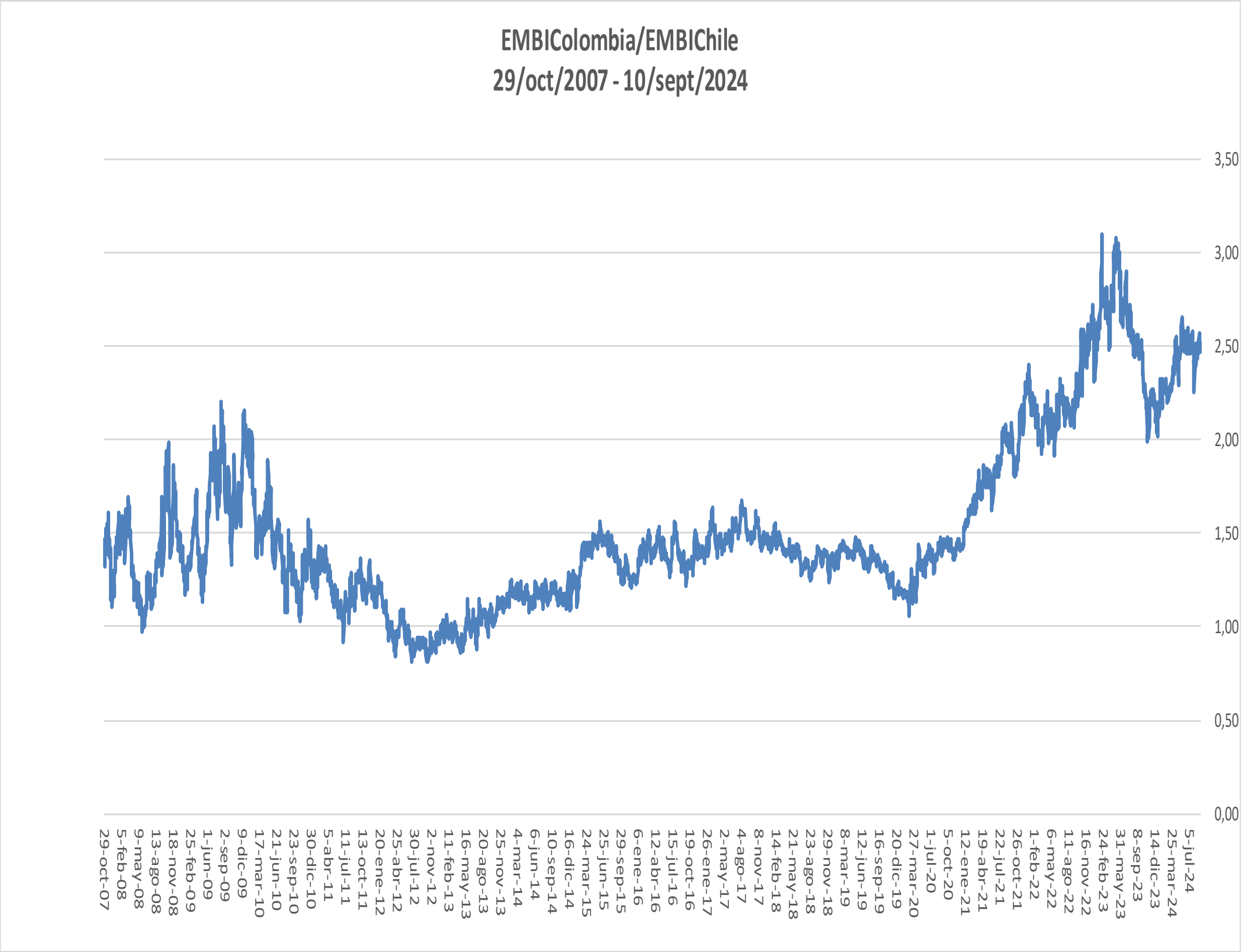

The EMBI indicator has been available since October 29, 2007. The one corresponding to Chile has been, in general, the lowest of those recorded for Latin American countries. The average of the ratio between the EMBI Colombia and the EMBI Chile for the first period (between October 29/07 and June 29/18) was 1.32, while the average for the following period (Jun 30/18) – sept10/24) was 1.88. The percentage variation between these averages was 42%. So, assuming then the Colombian uncertainty degree increases by 42%, how much is the social cost of that?

This cost can be measured as the loss resulting from the comparison between the present values of two alternative trajectories of the real gross domestic product estimated with a certain model (I describe it in V section) along two paths, each with a horizon of a certain number of years and differing only in the magnitude of the uncertainty degree (in one case this degree is low, 10%, and in the other it is high, 14.2%, that is, 42% higher).

For a 15-year horizon, the loss is 0.5% (equal to 1 minus the ratio between the present value of what is produced with the higher uncertainty and the present value of the product that could have been obtained under conditions of low uncertainty), assuming the innovation activity carried out by firms is not greatly affected by the greater uncertainty (i.e., assuming that the greater risk only reduces the capital/labor measured in efficiency units ratio) or 1.6% if the greater uncertainty has a greater negative effect, reducing the probability of successful results of the innovation activity. As will be obvious to the reader, the estimation is dependent on the characteristics of the model with which the GDP trajectories were generated in both alternatives.

## V. The model

The model I used to generate the GDP trajectories with low and high uncertainty is an adaptation of the A&H´s long-term growth model. I assumed the economy's savings rate has an endogenous component: that associated to a certain uncertainty degree.

Precisely, what was assumed, trying not to stray too far from the basic structure of the A&H´s model, was to establish this hypothesis: the savings rate ($s$: the savings/output ratio) is equal to an exogenous component multiplied by a variable that explicitly incorporates the degree of uncertainty, thus: $s = \bar{s}\tilde{k}^{(\pm)\eta}$, with $\bar{s}$ being the exogenous component of the savings rate (the only one extant in the A&H´s model), $\tilde{k}$ being the ratio of total capital to the variable *AN* (the multiple of the multifactor productivity factor, *A*, and the labor force, *N*), and $\eta$ being the degree of uncertainty ($\eta \geq 0$).

According to Ramey and Ramey (1995) the negative correlation between risk (I proxied it by volatility) and economic growth is higher (in absolute value and more reliable) for the subsample of less developed countries; and for the subsample of only developed countries (*OECD*) the resulting correlation was negative but lower (in absolute value) or null[10]. Therefore, considering the classification of economies I was made in this paper, it is best, in terms of the modified A&H´s model, to consider the following: "robust economies": $s = \bar{s}\tilde{k}^{\eta}$; "frail economies": $s = \bar{s}\tilde{k}^{(-)\eta}$.

In terms of Carroll *et al*. (2019), and for the case of the "robust economy", what this means is the following: given a target for future wealth (the "*target wealth"*), the higher the uncertainty the higher the savings rate with respect to the initial capital. Symmetrically, in the “frail economy” a greater uncertainty causes more capital outflow and so less capital invested there.

[10] A&H (chap. 14) sought to theoretically explain the results of Ramey and Ramey (1995) using two sorts of models: 1) AK models (the one is that of Jones *et al.*, 2000) in which higher volatility has a negative effect on growth mainly through the channel of higher risk on investment; 2) models in which higher volatility is associated with higher fluctuation of multifactor productivity which results in higher investment volatility and lower investment and GDP growth in the long run, because of credit constraints.

In all other respects, the model is the same as A&D´s model. The reduced form of the modified model is as follows (for a steady state path):

$$g_A = (\gamma - 1)\lambda\left[\alpha(1-\alpha)\lambda\sigma\tilde{k}^{\alpha}\right]^{\frac{\sigma}{1-\sigma}} \qquad (1)$$

$$\tilde{k} = \left(\frac{\bar{s}}{g_A + g_N + \delta}\right)^{\frac{1}{1-\alpha-(\pm)\eta}} \qquad (2)$$

$$r = \alpha^2\tilde{k}^{\alpha-1} - \delta \qquad (3)$$

The parameters are represented by Greek letters (in addition to $\bar{s}$), where $g_A, g_N, \delta, r$ are the multifactor productivity and labor growth rates, the depreciation rate and the real interest rate. The endogenous variables are $g_A, \tilde{k}, r$ for the steady state. And since in the steady state the GDP growth rate is the sum $g_A + g_N$ (considering that $\tilde{k}$ stabilizes at its steady state level), then the model also determines GDP growth rate on a steady state path.

The parameter $\gamma$ measures what would be the innovation rate under full certainty (which, by hypothesis, is never reached); $\lambda$ is a parameter helping to determine the probability of success of the innovative activity; notwithstanding, this activity requires investing capital (allocating financial resources to research and development of new products), and its efficiency also depends on another parameter, $\sigma$, which is the elasticity of the probability of success of the innovation with respect to the capital invested there. In addition, $\alpha$ is the elasticity-capital of the output.

Figure 3 illustrates the model´s reduced form.

Figure 3

**Impacts of higher precautionary savings in the “frail economy” (the A&H modified model)**

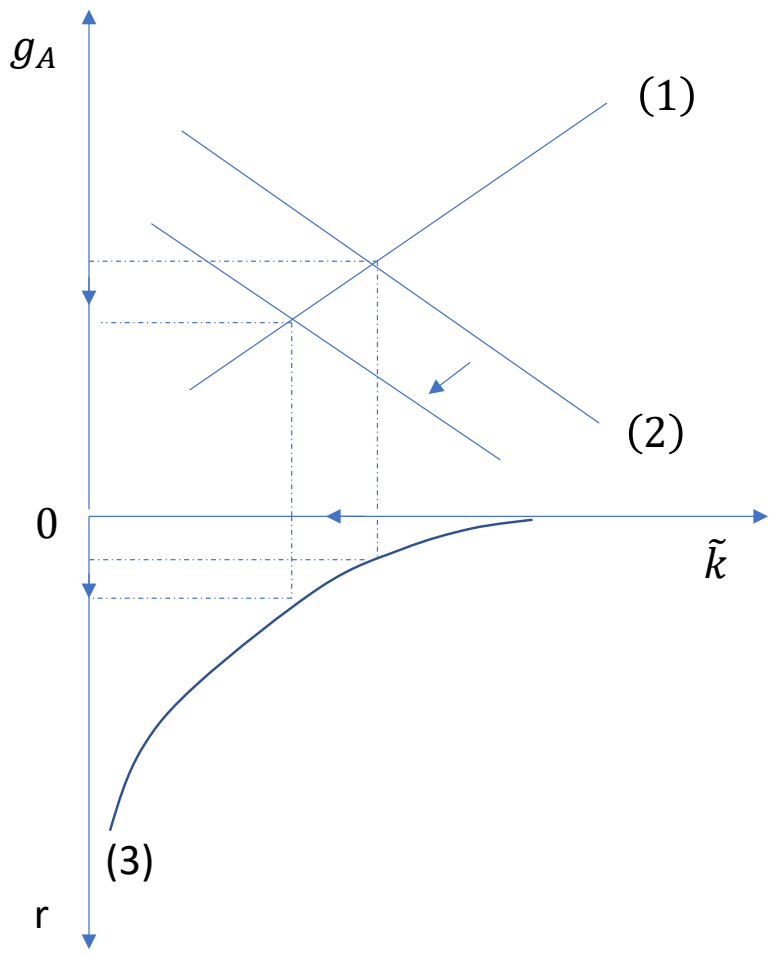

For the parameters, I assigned usual values in macroeconomic models or sought to avoid implausible results. But there is one, $\lambda$, whose magnitude has a particularly high incidence on the results, but it is not easy to find estimates for its value in the traditional literature or in statistical data. I ciphered the value of λ from repeated exercises of replication of probabilities of success in the innovation process with a probability law called “Pareto distribution” taking series of random numbers as inputs. I did this with the guidance in section 2 of Eaton and Kortum (2024).

Tables 1 and 2 show the parameter values and the main results of the model for the “open economy” case.

| **Table 2. Parameters** | $\gamma$ | $\lambda$ | $\alpha$ | $\sigma$ | $\eta$ | $\delta$ | $g_N$ | $\bar{s}$ |
|---|---|---|---|---|---|---|---|---|
| Baseline Scenario | 1.102 | 0.88 | 0.4 | 0.5 | -0.1 | 0.045 | 0.01 | 0.25 |
| Alternative Scenario 1 (higher risk) | = | = | = | = | -0.42 | = | = | = |
| Alternative Scenario 2 (higher risk and lower probability of innovation) | = | 0.84 | = | = | -0.42 | = | = | = |

| **Tabla 3. Model´s Output** | $\tilde{k}$ | $g_A$ | $g_Y = g_A + g_N$ | $r$ |
|---|---|---|---|---|
| Baseline Scenario | 5,69 | 0,019 | 0,029 | 0,0114 |
| Alternative Scenario 1 (higher risk) | 5,22 | 0,018 | 0,028 | 0,0144 |
| Alternative Scenario 2 (higher risk and lower probability of innovation) | 5,37 | 0,017 | 0,027 | 0,0134 |

There are two additional scenarios to the baseline scenario: the first (Scenario 1) contemplates a greater risk, and the second (Scenario 2) assumes that a greater risk goes hand in hand with a lower probability of success in the activities with which innovation is sought (this second scenario implies, according to Figure 3, a shift of curve 1 to the left). The results presented in the third column were used to calculate the social cost of uncertainty.

## VI. Conclusion

The main conclusion from all the above is the following: An increase in the degree of (bad) uncertainty does not cause a high social loss. The big loss, if any, would seem to arise from policy changes begetting negative effects on GDP on top of the uncertainty, and beyond what is caused by the increase in uncertainty itself.